\documentclass[aps,floats,showpacs,amsmath]{revtex4}
\usepackage{amssymb,amsmath}
\usepackage{graphicx,float}
\usepackage{epsfig}

\begin{document}
\title{Collective Charge Fluctuations in Single-Electron Processes on Nano-Networks}
\author{\large  \flushleft 
Milovan \v Suvakov and  Bosiljka Tadi\'c  
} 

\affiliation{ 
$^2$Department  for Theoretical Physics; Jo\v{z}ef Stefan Institute; 
P.O. Box 3000; SI-1001 Ljubljana; Slovenia.\\ \hspace{1cm}} 

\begin{abstract}
Using numerical modeling we study emergence of structure and structure-related nonlinear conduction  properties in the self-assembled nanoparticle films.
Particularly, we show how different nanoparticle networks  emerge within assembly processes with molecular bio-recognition binding. We then simulate the charge transport under voltage bias via single-electron tunnelings  through the junctions between nanoparticles on such type of networks. We show how the regular nanoparticle array and topologically inhomogeneous nanonetworks affect the charge transport. We find long-range correlations in the time series of charge fluctuation at individual nanoparticles and of flow along the junctions within the network. These correlations explain the occurrence of a large nonlinearity in the simulated and experimentally measured current-voltage characteristics and  non-Gaussian fluctuations of the current at the electrode.     
\end{abstract}
\maketitle
\section{Introduction}
Functional soft materials composed of many nano-particles often appear in  complex structure both as 2-dimensional films \cite{philip02,SA-films} and 3-dimensional super-crystal structures \cite{Mirkin,DNA08-1,DNA08-2}. These structures emerge in various {\it processes of self-assembly}, most of which  utilize liquid degrees of freedom  and  nonequilibrium methods \cite{Likos,Oxford}. 
The gold  nanoparticles are often used 
due to the property that their surfaces  can be {\it functionalized}  in an appropriate manner. The functionalization  of the nanoparticle surfaces has effects on their interactions \cite{NT-book}.
In the case of nanoaprticle films on substrate assembled by the liquid evaporation  \cite{Mathias} the 
thiol-passivated  nanoparticles where used, where the attached chains 
determine the minimum distance between two neighbouring nanoparticles.   
Another way to induce desired interaction between two nanoparticles was designed recently with attaching the pieces of biologically compatible strands of DNA molecules to different nanoparticles \cite{Mirkin}. The DNA-mediated interaction then utilizes the {\it biological recognition} and binding energy of the compatible DNA base pairs. Depending on the number of other factors, which we will also discuss later, this programmable self-assembly may lead to random as well as regular super-crystal structures, as shown in recent experiments in Refs.\  \cite{Mirkin,DNA08-1,DNA08-2}. Somewhat similar (but in a different potential  range),  is the bio-recognition assembly which utilizes biologically compatible protein molecules, for instance the antigen--antibody recognition \cite{AA}, which can be used in biosensors.

The  nanoparticle assemblies exhibit a wide range of physical properties which are not found in the bulk materials and at the level of a single nanoparticle \cite{NT-book}. Moreover, the emergent physical properties of the nanoparticle assemblies can often be related to their spatial structure \cite{pileni-review,Oxford,DNA08-1,DNA08-2}. For instance, new optical and magnetic properties have been found in super-lattices made of nanocrystal \cite{pileni-review}. By  speciffic functionalization of their surfaces gold nanoparticles develop directionally dependent  exchange interaction \cite{exchange}. 
Assuming that such nanoparticles can be arranged on nodes of  a sparse scale-free graph,  numerical simulations predict new features of the magnetic response of such a structure \cite{BT-book,PRL}, compared to the classical memory materials \cite{PhysA}. Similarly, the films of gold nanoparticles  under voltage bias conduct current via {\it single-electron tunneling} processes \cite{SET} due to their small capacitances and the Coulomb blocade effects \cite{CB-book}.  In contrast to classical conduction, the single-electron conduction processes  have numerous advantages, e.g., the absense of dissipation due to  heathing, and the nonlinear current-voltage characteristics, which are important in the nanoelectronics. Experimentally a large nonlinearity of the current--voltage curve has been found in self-assembled nanoparticle films on substrates, which appears to be fairly correlated with the film structure \cite{parthasarathy01,philip02,NL07}.

The random structure of nanoparticle films on substrates can be adequately modelled by planar graphs ({\it nano-networks}) \cite{LNCS06,NL07}. For the charge transport via single-electron processes, which we study here, the topological elements of these nano-networks are suitably selected as follows: the  {\it nodes} represent nanoparticles, whereas {\it edges} (links) are the tunneling junctions between two nanoparticles. For the junction (link) to occur between two nanoparticles,  the  distance between them must be smaller than the {\it tunneling radius} of the electrons for that kind of nanoparticles \cite{CB-book}. 
The charge transport through the film under applied voltage can be described within the  model of the capacitively-coupled nanoparticles  on substrate \cite{Middleton}. In our previous work \cite{ST08} we generalized the model of capacitively-coupled nanoparticle assemblies for an arbitrary topology of the nano-networks \cite{NL07,ST08} and implemented a numerical model of single-electron tunnelings driven by the external voltage through such networks. Theoretical background and the details of the numerical implementation is reviewed in Ref.\  \cite{ST08}. The simulation results revealed the topology-induced nonlinearity in the $I(V)$ curve, in  a good agreement with the experiments \cite{NL07}.
 
In the present work we study the collective dynamic behavior which makes the physical basis  for the observed $I(V)$  nonlinearity, in particular, the long-range correlations in charge fluctuations monitored at each nanoparticle and inhomogeneous charge flow along the conducting paths through the sample and at the electrode.  We determine the probability distributions  of various quantities related to the conduction process and demonstrate 
how the topological and charge disorder affect the quantitative statistical properties and the $I(V)$ dependences.

The organization of the paper is as follows: In Section\ 2 we present the model of the particle self-assembly with bio-recognition bonding and show some of the typical emergent structures. In section\ 3 we summarize  the theoretical background of the conduction with single-electron processes and its numerical implementation for an arbitrary network topology. The visualization of the conduction paths through two network types, the regular and  inhomogeneous nanoparticle aggregates, as well as the statistical properties of the topological and charge flow are shown in Section\ 4. In Section \ 5 we analyse the simulated time series of charge and current fluctuations and the related $I(V)$ curves. Section \ 6 gives a short summary of the results.

\section{ Nanoparticle networks emerging in self-assembly with bio-recognition}
Different assembly processes make use of the liquid phase, that provides the nanoparticle diffusion and affects the interaction. The assembly with fast  evaporation of the liquid \cite{Mathias,philip02} induces additional constraints to the particle motion and interaction.
The interaction itself between pairs of the nanoparticles moving in a liquid is taking part at short distances and depends on the type of functionalization of their surfaces, as mentioned above. 
In the case of bio-recognition binding, which we will discuss shortly here, the attached  DNA induce the differences between two types of nanoparticles, i.e., type A and type B nanoparticles, each of which carry the biologically compatible strand of the same DNA molecule. Thus, the interaction between A--B type is attractive within some bonding distance $d$. The bonding distance depends on  the lengths of the attached molecules and on the number of the compatible base-pairs on them \cite{DNA08-1}. Therefore, it exceeds the sum of the particle radii $d_b> R_A+R_B$. The effective interaction between A--B particles  can be then modeled by the Lennard-Jones potential \cite{colloids-numerics}
\begin{equation}
V_{A-B}(r)=4\epsilon \left( \frac{\sigma^{12}}{(r-R_A-R_B)^{12}} - \frac{\sigma^{6}}{(r-R_A-R_B)^{6}} \right),
\label{eq-LJpotential}
\end{equation}
where $r$ is the actual distance between particles and the potential minimum corresponds to the optimal bonding distance. In the present case, the parameters $\sigma$ and $\epsilon$ depend, roughly speaking, on the lengths of the attached DNA chains and the strength of binding (number of the compatible base pairs)\cite{Oxford}. On the contrary, the particles which carry biologically incompatible DNA strands have no attractive interaction of this type. The repulsive part alone of the potential in Eq.\ (\ref{eq-LJpotential}) can describe the interaction of the equal-type particles A--A and B--B, occurring at short distances. 

\begin{figure}[htb]
\begin{tabular}{cc} 
\resizebox{18.4pc}{!}{\includegraphics{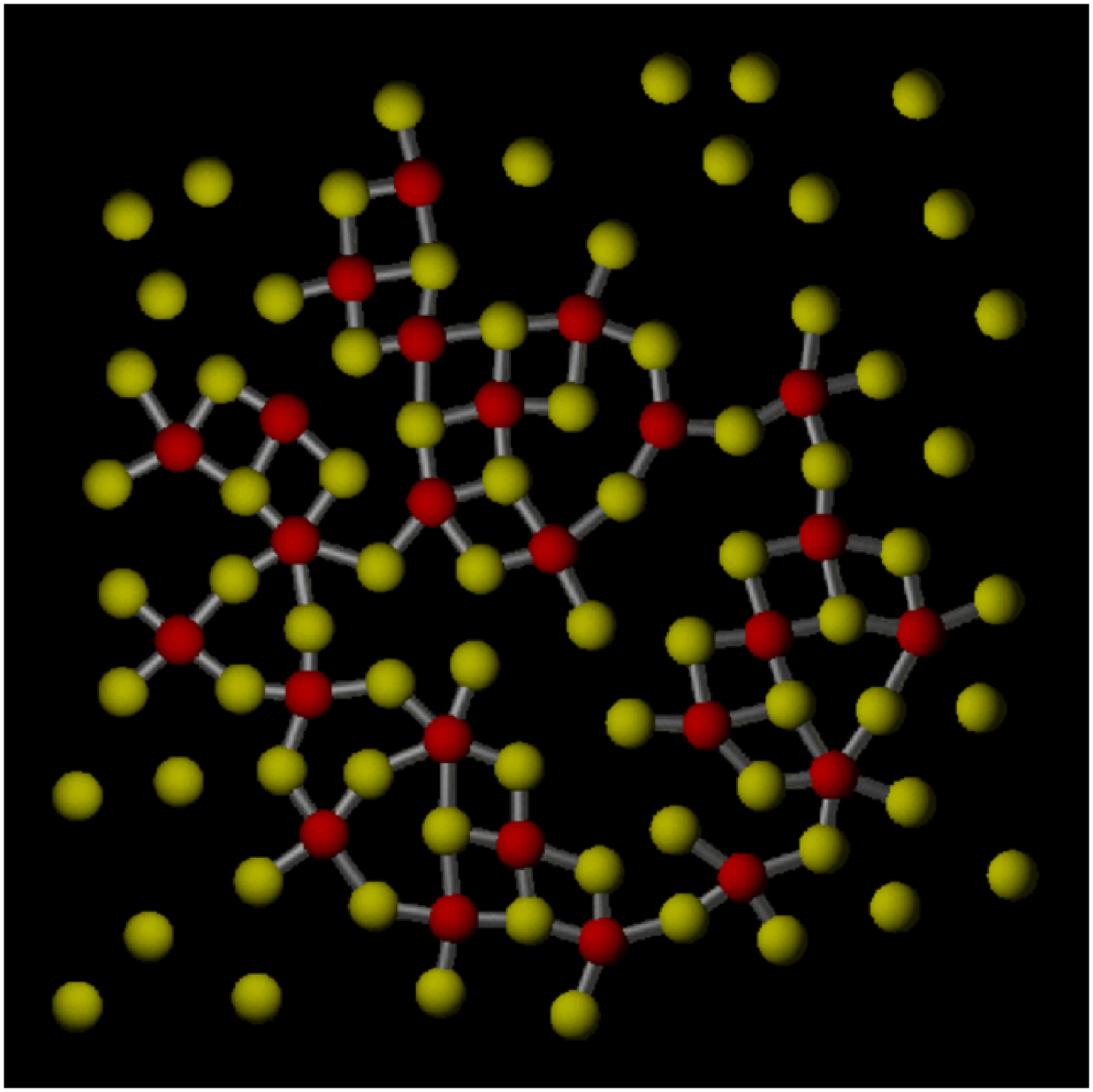}}&
\resizebox{18.4pc}{!}{\includegraphics{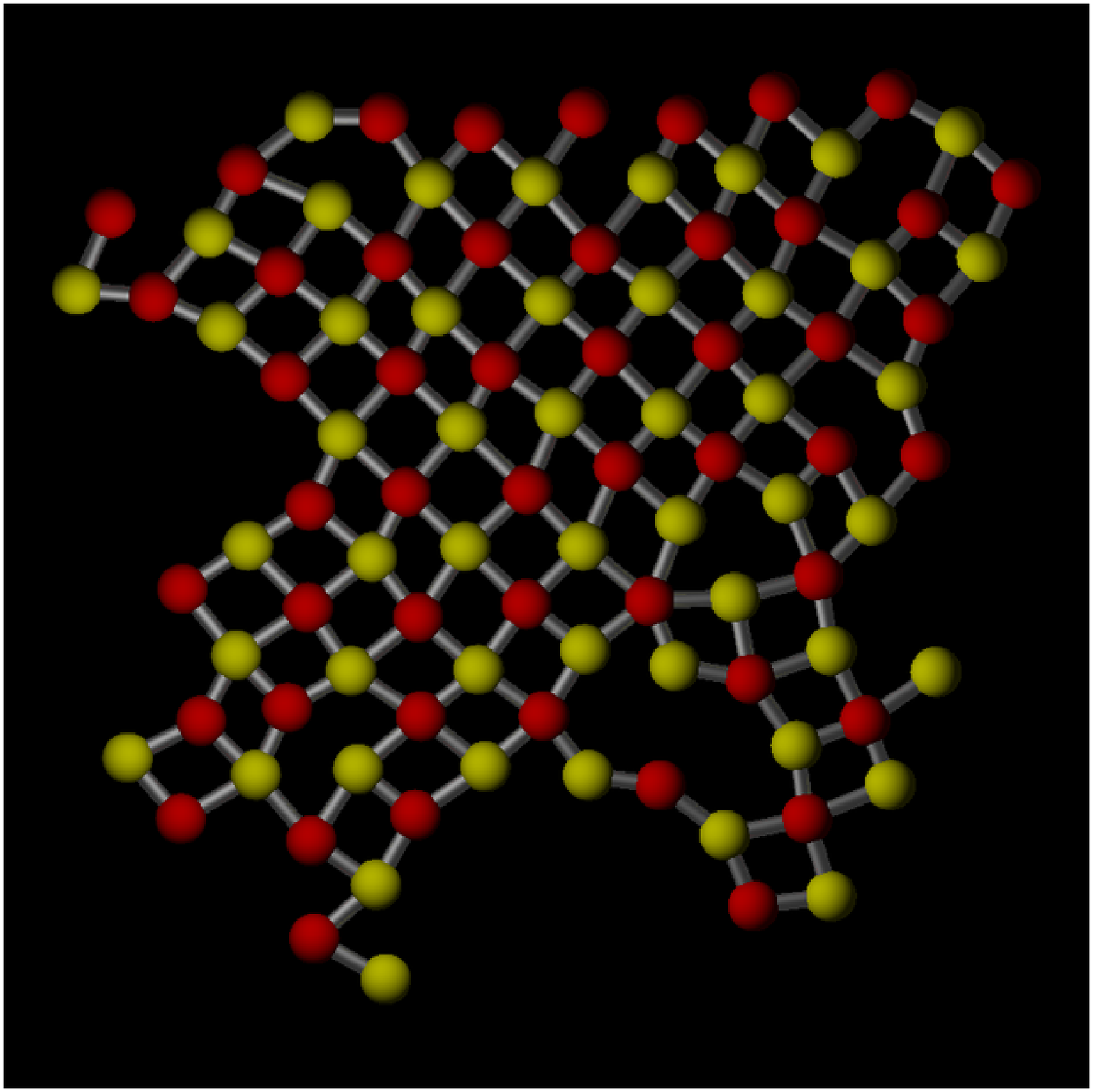}}\\
\end{tabular}
\caption{Two examples of the emergent nanoparticle films obtained via bio-recognition binding between A (red) and B (yellow) particles for equal coverage and different relative concentrations $c_A/c_B$ as follows 25\% : 75\% (left) and 50\% : 50\% (right).}
\label{fig-twostructures}
\end{figure}
The diffusion of such nanoparticles can be simulated via the Langevin equation with the random force originating from the integrated liquid degrees of freedom and the thermal fluctuations \cite{ST-colloids}. Apart form the molecular difference between attached DNA strands, generally, the particle radii $R_A \neq R_B$ can be different and consequently two different diffusion coefficient enter the equation
\begin{equation}\label{eqm}
\nu_i {\dot {\mathbf r}_i}=-\nabla_ {{\mathbf r} _i} 
\sum_{j}V_{i-j}(\vert {\mathbf r} _i - {\mathbf r} _j\vert)+F_{i,T},
\end{equation}
with $
 \left\langle F_{i,T}(t) \right\rangle = 0$,  
 $\left\langle F_{i,T} F_{j,T} (t') \right\rangle = 6k_BT\nu_i\delta_{i,j}\delta_{t,t'}$,  and 
$D_i=k_BT/\nu_i$, $i$ referring to either A or B particle type.
A complete description of the model and the discussion of the role of different parameters in the self-assembly processes will be given elsewhere \cite{ST-colloids}. Here we restrict our discussion to the situation which is interesting for the assembly of the conducting nanoparticle films.  Therefore, we consider particles of equal size $R_A=R_B$, and vary the relative concentration $c_A:c_B$ of particles A and B carrying with different strands of the molecule. In this way various structures can be assembled, both regular and irregular, which in turn appear to have different physical properties (as discussed below).  

In Fig.\ \ref{fig-twostructures} we show two representative structures which emerge in the process when temperature is kept low compared to the parameter $\epsilon$, and the total coverage of the surface by particles is fixed, but the relative concentrations of A and B particles varied. Generally, for low enough temperatures and equal concentrations the system has tendency towards a {\it regular} ordered structure with few defects (Fig.\ \ref{fig-twostructures}, right). Whereas, when the concentrations are in disballance, another type of structure with local order and large voids may appear, as shown in Fig.\ \ref{fig-twostructures}, left. It should be stressed that, for a fixed set of parameters, many such structures may appear depending on the initial conditions. Each of them corresponds to a local free energy minimum, and appear to be quite stable for the low temperatures \cite{ST-colloids}.
In the following we will discuss in detail the conduction properties of a regular nanoparticle structure and of a highly irregular nanoparticle structure on the substrate. In practice, the regular (triangular) nanoparticle arrays \cite{parthasarathy01} and irregular nanoparticle films \cite{SA-films} are assembled using entirely different approaches. 
Our results shown above demonstrate that such diverse structures may appear in the same self-assembly process with bio-recognition when a single  parameter (the relative concentration of differently functionalized nanoparticles) is suitably varied.

\section{Single-electron tunnelings in nanoparticle films}
The capacitively coupled nanoparticle assembly on substrate under applied voltage conducts current with a cascade processes of single-electron tunnelings, which is enabled by the Coulomb blockade  at junctions between the adjacent nanoparticles \cite{CB-book,Middleton}. Following a single electron tunneling through the junction between two particles $a\to b$, the charging energy of the nanoparticle increases inversely proportional to its capacitance. This increases the barrier for further tunnelings along that junction, and consequently, no tunnelings occur until the voltage is increased enough to overcome the barrier.

The theoretical background of the tunnelings within the nanoparticle arrays has been developed in \cite{Middleton,parthasarathy04}. Recently we have generalized the formulation for the case of nano-networks with the tunneling junctions representing as the network edges \cite{ST08}. It should be stressed that the links in the nano-network suitable for the tunneling processes might differ form the ones of the bio-recognition binding, shown in Fig.\ \ref{fig-twostructures}. Specifically, for the tunneling processes important 
are positions of the nanoparticles on the substrate and their mutual distances compared to the tunneling radius $r_t$.  The chemical bonds, which are  shown in Fig.\ \ref{fig-twostructures}, are to be understood as fixing the nanoparticle positions. Then the tunneling junction network may have the links which coincide with the chemical bonds only when the bonding distance $d_b\sim r_t$, otherwise, the tunneling junction network is different. Generally, the nano-network 
 edge representing the tunneling junction occurs between any two nanoparticles which are separated at distances $r\leq r_t$. In addition, the difference between A and B particle types, which was important for the assembly, plays no role in the conduction (assuming the particles of the same size and  chemical composition). In this way, a nano-network of tunneling junctions can be constructed whenever the positions of the nanoparticles on the substrate are known \cite{LNCS07}. In our numerical models the tunneling radius is a free parameter. Varying the tunneling radius, e.g., by changing the chemical composition and size of the nanoparticles, slightly different nano-networks of junctions can be constructed. For the numerical implementation of the tunneling process the nano-network of tunneling junctions is fixed by the adjacency matrix with the elemnts $A_{ij}=1$, when the tunneling between the nanoparticles $i$ and $j$ can occur, and $ A_{ij}=0$, when the junction is absent, e.g., due to large distance $d_{ij} >r_t$ between them.

\subsection{Electrostatic energy of a nanoparticle network}
Following the classical work \cite{Middleton} the electrostatic energy for a capacitively coupled nanoparticle array consisting of $N$ nanoparticles and of a general topology given by the adjacency matrix $\{A_{ij}\}$ can be  described as follows \cite{ST08}.
The capacitances are related to the structure of the underlying nano-network by $C_{ij}\equiv CA_{ij}$.  The charge  $Q_i$  on $i$th nanoparticle can be expressed as 
\begin{equation}
  Q_i=\sum_j C_{ij}(\Phi_i-\Phi_j) + \sum_\mu C_{i,\mu}(\Phi_i-\Phi_\mu) ,
\label{charge}
\end{equation}
where $\Phi_i$ is the potential of the $i$th nanoparticle in the assembly. 
The index $\mu \in \lbrace +,-,g\rbrace$ is attributed to the electrodes (including  the gate) and  thus $\Phi_\mu$ is the potential at the $\mu$th electrode and  similarly $Q_{\mu}$ is the charge and  $C_{i,\mu}$ the capacitance between the electrode $\mu$th electrode  and $i$th the nanoparticle.
The system of equations (\ref{charge}) for $i=1,2, \cdots N$ can be written in the matrix form 
 $\mathbf{Q} = M\mathbf{\Phi} - \mathbf{C}_\mu \Phi^{\mu}$ and formally solved for $\mathbf{\Phi}$  as

\begin{equation}
 \Phi = M^{-1} \mathbf{Q} + M^{-1} \mathbf{C}_\mu \Phi^{\mu} \ 
\label{phi}
\end{equation}
where now $\mathbf{M}$ is the capacitance matrix of the whole system (nanoparticles plus electrodes) with the elements:
\begin{equation}
M_{ij}=\delta_{i,j}(\sum_k C_{ik}+\sum_\mu C_{i,\mu})-C_{ij} \ .
\label{Mmatrix}
\end{equation}
Using these definitions, the electrostatic energy of the whole system can be written as

\begin{equation}
E=\frac{1}{2} \mathbf{Q}^\dag M^{-1} \mathbf{Q} +  \mathbf{Q} \cdot V^{ext}  + Q_{\mu} \Phi^{\mu},\quad V^{ext}=M^{-1} \mathbf{C}_\mu \Phi^{\mu} .
\label{energy}
\end{equation}
where  $\mu \in \lbrace +,-,g\rbrace$ stands for the electrodes and gate, $V^{ext}$ is the external voltage and  $C_{i,\mu}$, $Q_{\mu}$, $\Phi_\mu$ are described above.

\subsection{Numerical implementation and visualization of charge transport}
The system is driven by increasing the external voltage at one, say left, electrode. Voltage on the gate is kept fixed. Due to the voltage bias, the electrons first tunnel from the left electrode to the neighboring nanoparticles and then forward along the available junctions between the nanoparticles. Due to the overall voltage profile the tunnelings are predominantly to the right, towards the zero-voltage electrode as $+V \to a \to b \to \cdots \to 0$. 

Following the  tunneling of a single electron from particle $a \to b$, causing the change in the particle charges as  
 $Q_i^\prime = Q_i + \delta_{ib} - \delta_{ia}$, the energy change $\Delta E (a \rightarrow b)$ occurs in the whole array, due the long-range interactions. 
The energy change is computed using the expression in Eq.\ (\ref{energy}) as follows:
\begin{equation}
\Delta E (a \rightarrow b) = V_b - V_a +  \frac{1}{2} (M^{-1}_{aa}+M^{-1}_{bb}-M^{-1}_{ab}-M^{-1}_{ba}) +V_b^{ext}-V_a^{ext}
\label{delta-energy}
\end{equation}
and similarly $\delta E(a \leftrightarrow \pm) = \pm V_a + \frac{1}{2} M^{-1}_{aa}$ originating from the tunneling between the electrode and the nanoparticle. 
Here we have defined the following quantity 
 \begin{equation}
V_c\equiv \sum_iQ_iM_{ic}^{-1} ,
\label{Vc}
\end{equation}
 which is also suitable for the numerical implementation. Namely, the updated value of the variable $V_c$ at node $c$ following the tunneling $a \rightarrow b$ can be computed recursively as follows:
\begin{equation}
 a \rightarrow b: \quad V'_c = V_c + M^{-1}_{bc} - M^{-1}_{ac}, ~~~ 
a \leftrightarrow \pm: \quad V'_c = V_c \pm M^{-1}_{ac}.
\label{Vcprime}
\end{equation}
The simulation takes the following steps: For the fixed voltage $V$ the whole network is updated for tunnelings. For each junction the attempted tunneling from $i \to j$ induces the global energy change $\Delta E{i\rightarrow j}$, which is computed using the above formulas. Then the  {\it tunneling rate} 
for $i \rightarrow j$ event at given voltage $V$ is computed according to \cite{ST08}
\begin{equation}
\Gamma_{i\rightarrow j}(V) = \frac{1}{eR_{i\rightarrow j}} \frac{\Delta E_{i\rightarrow j}/e}{1-e^{-\Delta E_{i\rightarrow j}/k_BT}}
\label{gammaij}
\end{equation}
where $e$ stands for the electron charge, $T$ is the temperature  and $R_{i\rightarrow j}$ is the {\it quantum resistance} of the junction. The quantum resistance (or conductance) can be computed for small nanoparticles. e.g., by density functional theory \cite{DFT}. For instance, for the metallic nanoparticles one can obtain  within the transfer Hamiltonian that $R_t=\frac{\hbar}{2\pi e^2 \left\vert T \right\vert^2 D_{l0} D_{r0}}$, where  $T_0$ and $D_0$  are  the transmission coefficient and the density of states at each nanoparticle \cite{CB-book}. We assume that  the quantum resistance$ R_{i\rightarrow j} \equiv R_t$ for all junctions. Note that in the approach described above, $R_t$ appears as a parameter which sets the characteristic time scale of the process via $\tau \sim 1/R_tC_g$ (see later).  
In our continuous time implementation, out of all attempted tunnelings only those which tunneling times $t_{i\rightarrow j}$ taken from the probability distribution $p(t_{i\rightarrow j}) = p_0exp(-\Gamma_{i\rightarrow j}t)$ appear to be in  small time window $\delta t$ are processed. Then the network is updated again (for more details see Ref.\ \cite{ST08}).Here we use the limit $C/C_g \ll 1$ and $V_g=0$.

\begin{figure}[htb] 
\resizebox{34pc}{24pc}{\includegraphics{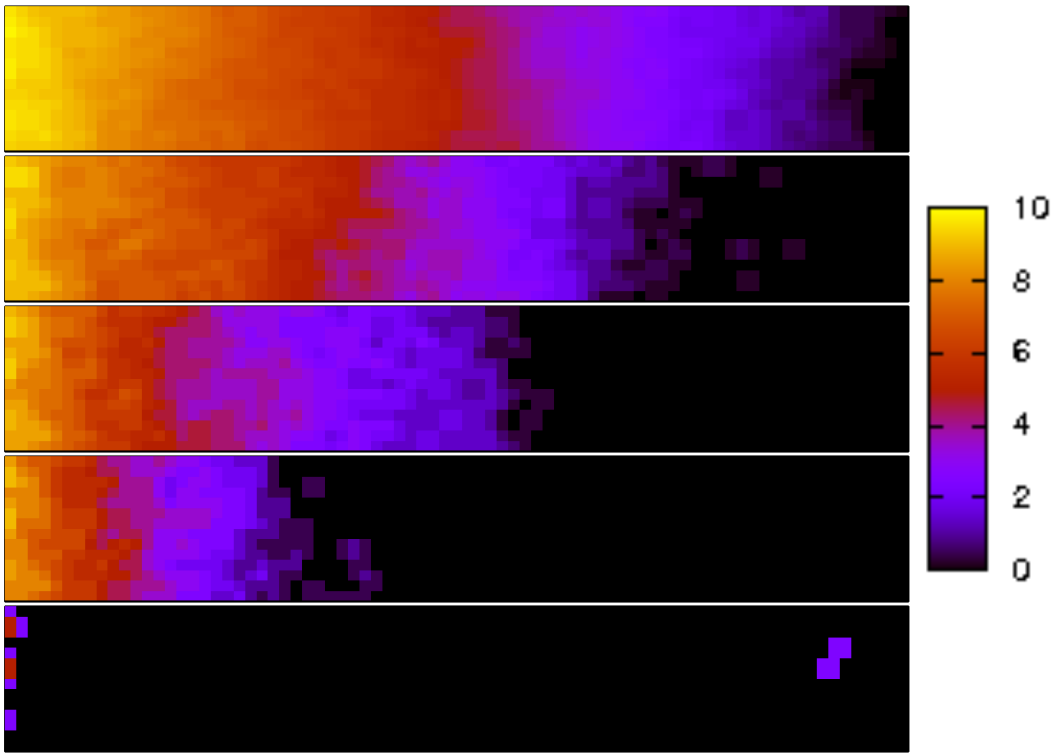}}
\label{figfront}
\caption{Visualization of the charge profile before the voltage threshold is reached.}
\end{figure}

When the external voltage is increased, more electrons tunnel into the nanoparticle array and consequently local ratio between the charge and voltage is changed, providing the conditions for more forward tunnelings. The moving front of charges on the slab of regular triangular arrays of nanoparticles is shown in Fig. \ref{figfront} for a set of different voltages. The color intensity indicates the number of electrones at the nanoparticle. When the voltage is large enough $V\sim V_T$, the moving front reaches the zero-voltage electrode and the current through the sample can be detected for the first time. 
For the voltages $V> V_T$  the current  through the sample is nonzero for all times. The average current  through the junction ${i\rightarrow j}$ is defined as the net number of electrons per time unit:  
\begin{equation} 
I_{i\leftrightarrow j}(V) = e\left(\Gamma _{i\rightarrow j}(V) - \Gamma _{j \rightarrow i}(V) \right)\ .
\label{currentij} 
 \end{equation}
In the numerical experiment we can measure the current through each junction within the film and at the electrode. In the laboratory experiments the current is measured only at the electrode. The current increases with increased voltage. However, the curren--voltage dependence appears to be non-linear for a range of voltages above the threshold $V > V_T$:
  \begin{equation}
I(V) = B \left({{V-V_T}\over{V_T}}\right)^\zeta \ ; \zeta > 1 \ ;
\label{eq-IV-zeta}
\end{equation}
The exponent $\zeta $ was found to vary from 2 to 4 with the structure of the nanoparticle arrays, in particular $\zeta =2.25$ was measured in the case of regular triangular array of gold nanoparticles \cite{parthasarathy04}, whereas $\zeta =3.9$ and larger was found in highly inhomogeneous films obtained by the evaporation methods \cite{NL07}.  As mentioned in the introduction, one of the purposes of the present  work is to study the charge and current fluctuations {\it inside the sample} in order to reveal the correlations in the process which contribute the observed nonlinearity in the current--voltage curves. The other aspect is to demonstrate the role of random charge disorder in the transport process. The charge disorder represents  fractional charges which are locally present in the real samples, for instance, due to differences in the contacts of some  nanoparticles with the substrate, and  affect the charge--voltage relation at  these nanoparticles.  In the simulations we consider the regular triangular array of nanoparticles with quenched charge disorder $0<q_i <1$ at $i$th nanoparticle taken from the uniform random distribution. The calculations are systematically compared with the case of nanoparticle film without charge disorder but with large topological inhomogeneity of the structure (Fig.\ \ref{fig-cpaths}).

\section{Topological and Charge flow: Conducting Paths}
\begin{figure}[htb]
\begin{tabular}{cc} 
\resizebox{16pc}{16pc}{\includegraphics{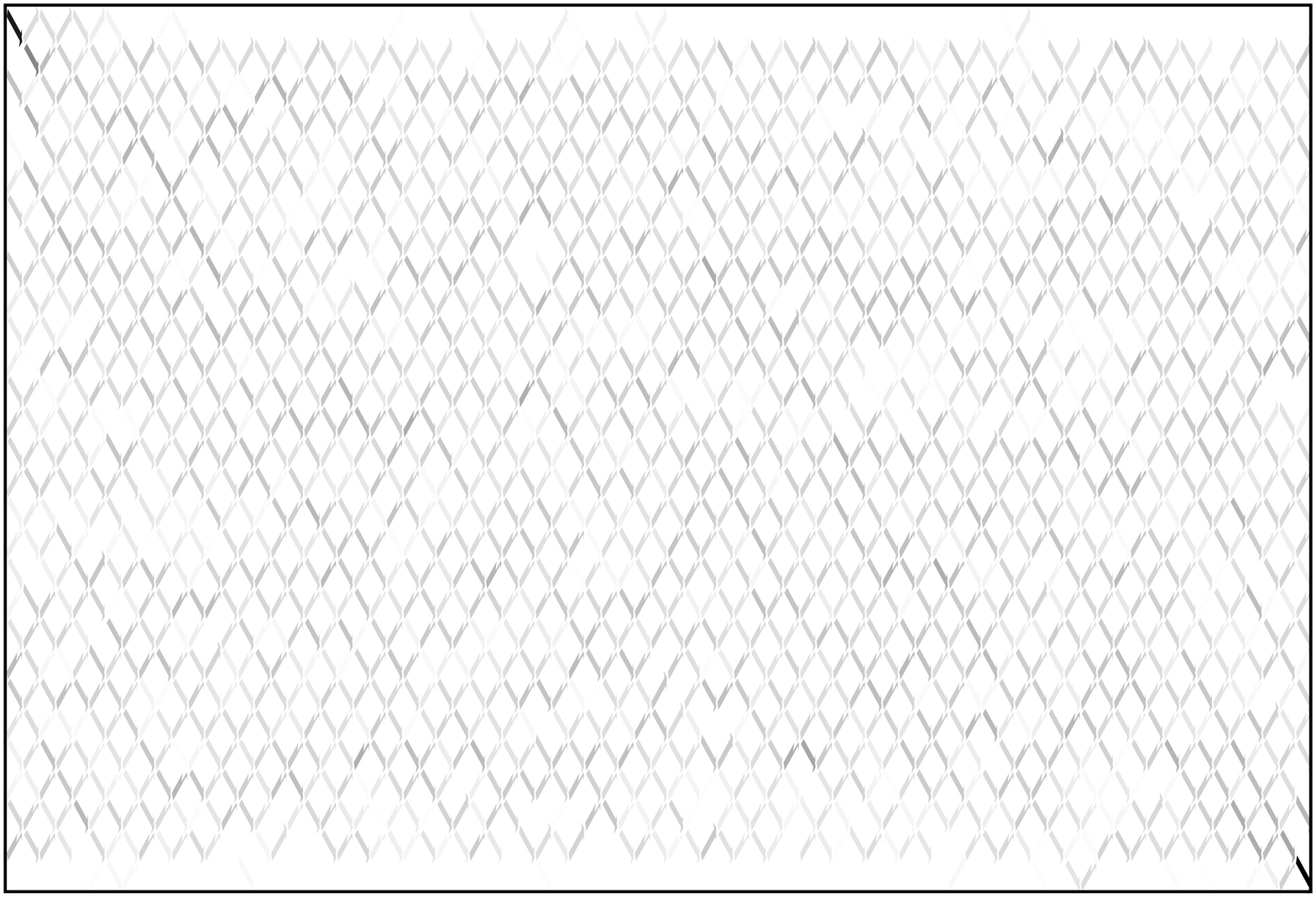}}&
\resizebox{16pc}{16pc}{\includegraphics{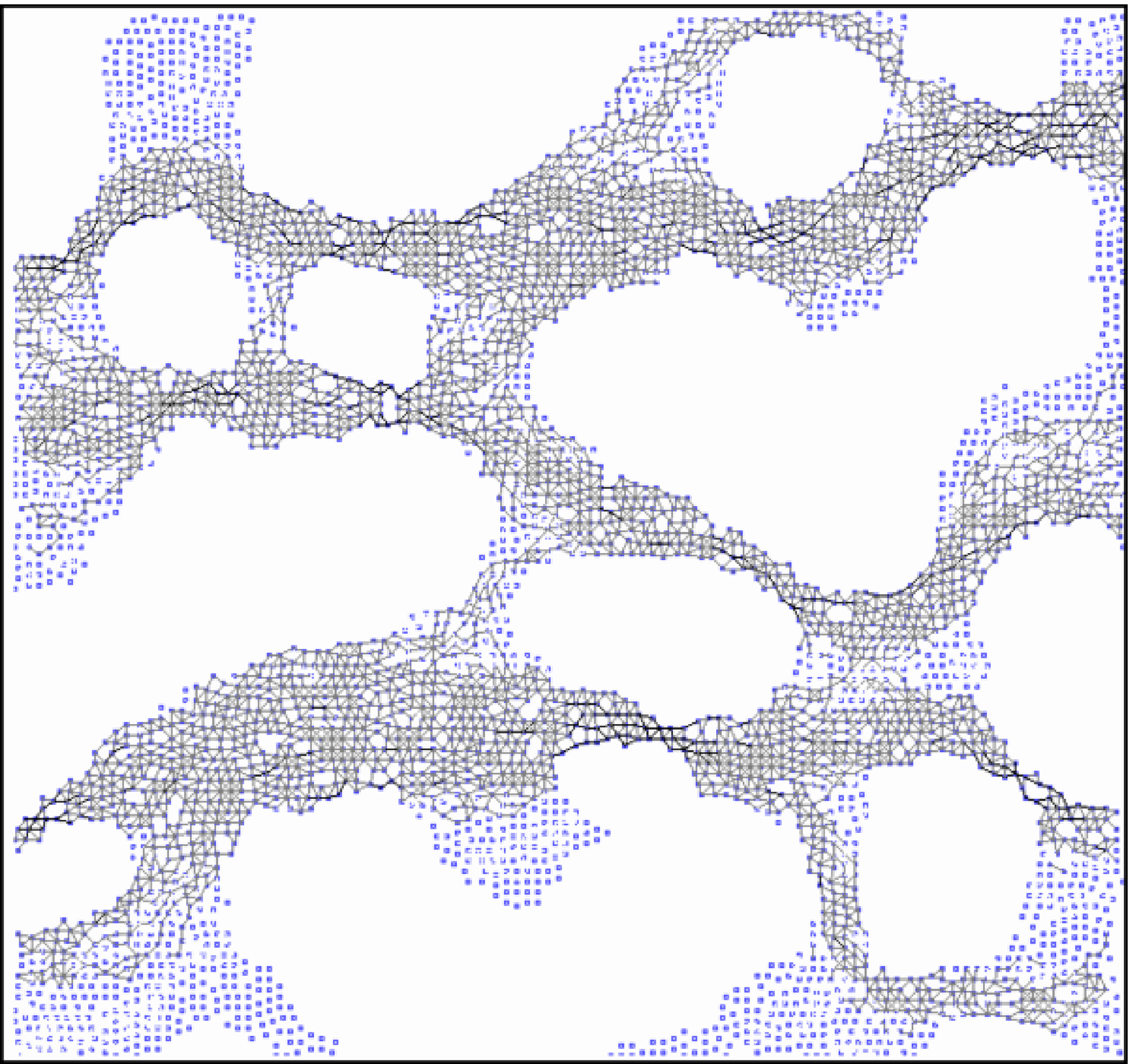}}\\
\end{tabular}\label{fig-cpaths}
\caption{Visualization of the conduction paths through the regular triangular array with charge disorder at nodes (left) and through the topologically inhomogeneous network without charge disorder (right). For a fixed voltage $V> V_T$ the cumulative charge flow along each link is recorded and represented by the color intensity.}
\end{figure}

\begin{figure}[htb]
\begin{tabular}{cc} 
\resizebox{18.4pc}{!}{\includegraphics{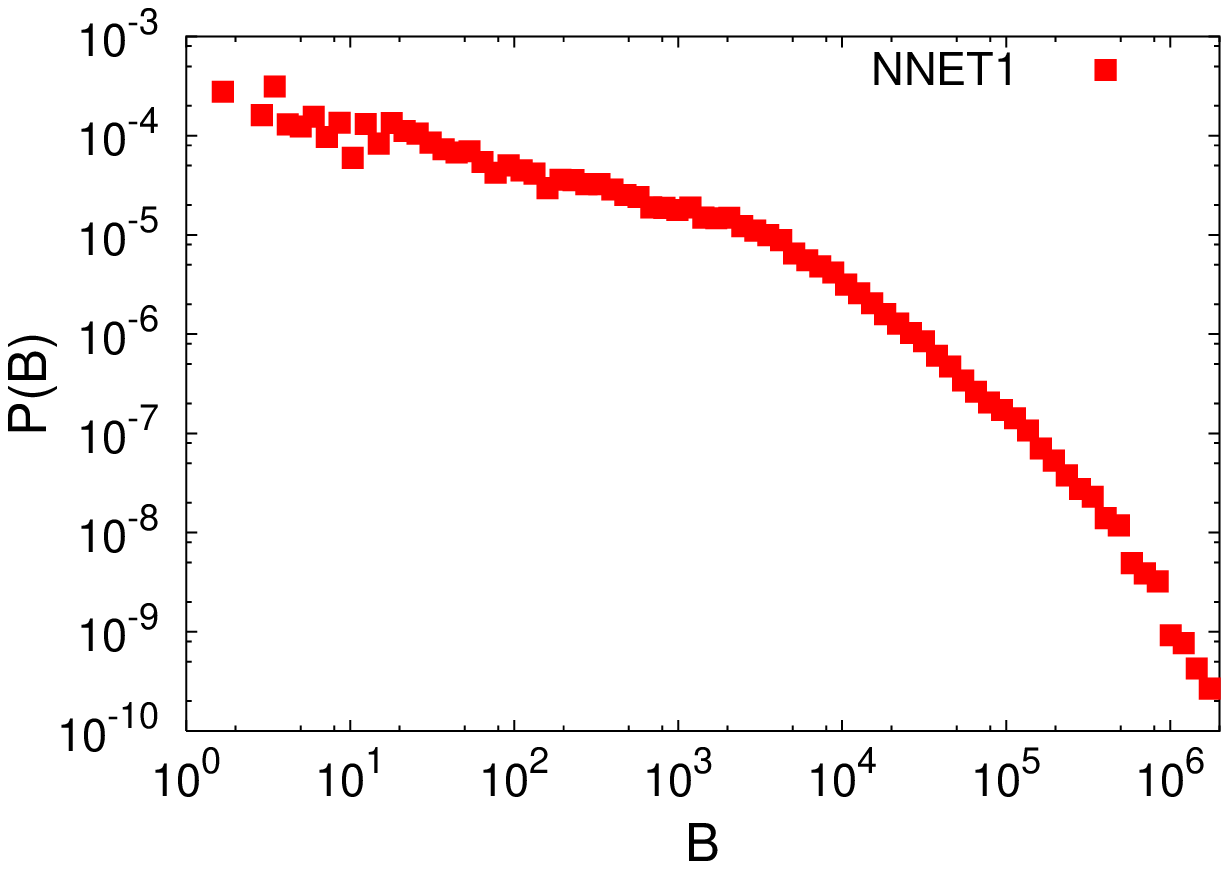}}&
\resizebox{18.4pc}{!}{\includegraphics{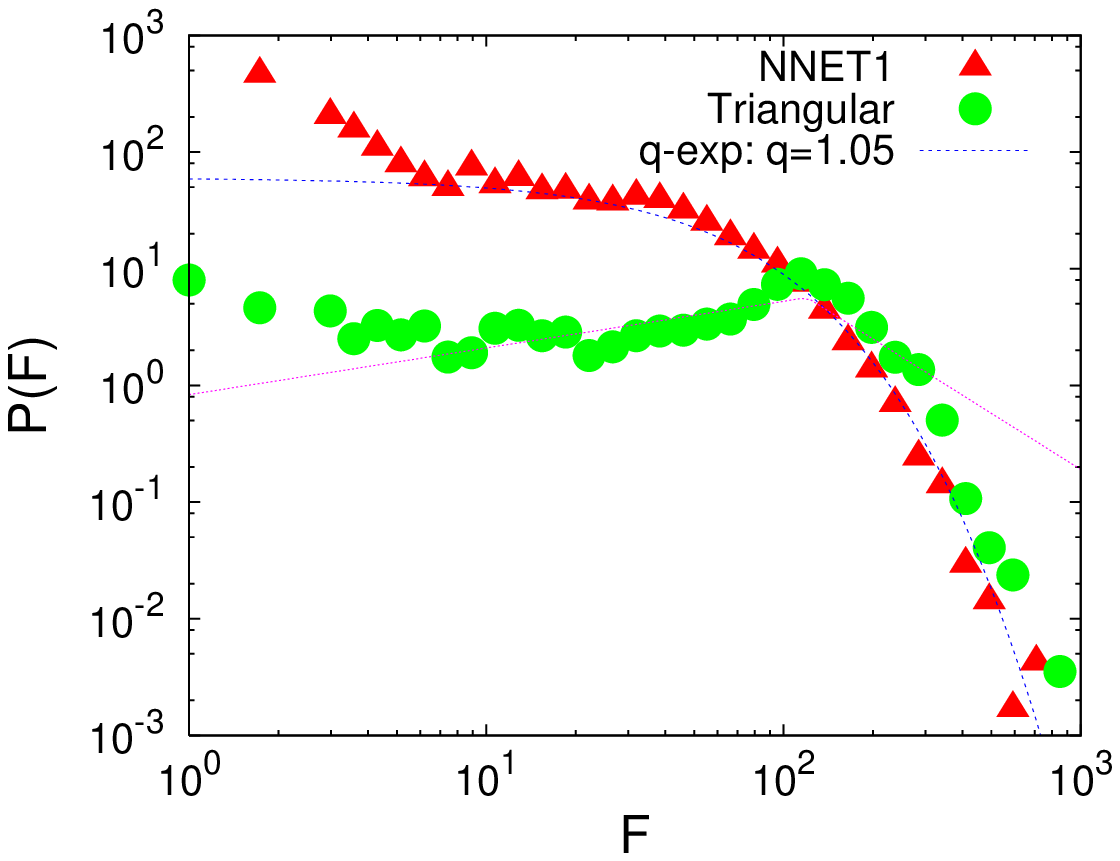}}\\
\end{tabular}
\caption{Distribution of the topological flow $B$ in the inhomogeneous nanonetwork  NNET1 (left) and distributions of charge flow $F$ along the links in  NNET1 and in the triangular nanoparticle array with charge disorder (right).}
\label{fig-BCs}
\end{figure}

For voltages over the threshold $V_T$ the {\it conducting paths} through the sample are established. By recording the current through  all junctions in the network we can visualize the charge flow along these conducting paths, as shown in Fig.\ \ref{fig-cpaths}. Darker color corresponds to larger number of electrons transported along each link. The Fig.\ \ref{fig-cpaths} also demonstrates the role of topological inhomogeneity, i.e., in the case of the nano-network in the right panel, as opposed to the regular nanoparticle array. In the inhomogeneous nano-network, the areas without nanoparticles make a restriction to the charge transport, thus the number of conducting paths merge into a smaller number in the bottleneck regions. Draining the electrons along such paths thus affects a large regions behand. This also indicates that certain nodes on the paths  play more important role in the charge flow than the others. The quantitative measure of the node importance is their topological betweenness centrality $B$, which is defined as the number of shortest paths between all pairs of nodes on the graph that go through that node. In the case of the inhomogeneous film network in Fig.\ \ref{fig-cpaths}(right),  the topological betweenness-centrality is given by a broad distribution with a power-law tail, shown in Fig.\ \ref{fig-BCs}(left). 
In the regular nano-array, on the contrary, the topological betweenness-centrality of all nodes is similar (not shown), which induces a less intricate pattern of conducting paths. In Fig.\ \ref{fig-cpaths}(left) we show the conducting paths in the case of a triangular array in the presence of charge disorder. 
As already mentioned above, the presence of a random fractional charge at nanoparticle may destroy the local balance between the voltage and charge, leading to blocking of certain conducting paths through that nanoparticle and consequently reduced charge transport through that particle.  Such areas with reduced charge flow appear as bright spots in Fig.\ \ref{fig-cpaths}(left).   
The distributions of the actual charge flow (dynamical betweenness-centrality) for both networks is shown in Fig.\ \ref{fig-BCs}(right). Again the differences between two topologies play the role. In the inhomogeneous network NNET1, the distribution of the charge flow along the links on this network resembles the distribution of topological betweenness (one should stress that the shortest paths between the electrodes play the role in the conduction).  The simulation results can be approximately fitted with the $q$-exponential form \cite{Tsallis}  $ P(F) \sim [1-(1-q)(F/F_0)]^{1/(1-q)}$  with $q=1.05$. In the case of the triangular array with charge disorder, the distribution away from the peak area  indicates that certain links in the network curry less charge during the conduction process. The curve $ P(F) \sim (1/F)exp |ln (F/F_0)|$  approximates the numerical results  of the distribution near the peak area. 

\section{Non-Gaussian Current \& Charge Fluctuations}
In our simulations the current recorded at the right electrode for the applied voltages above the threshold $V> V_T$ can be compared with the experimental measurements in the nanoparticle films \cite{parthasarathy04,NL07}. The threshold $V_T$ depends on the network size (distance between the electrodes) and its structure.  

\subsection{Current--Voltage Curves}
\begin{figure}[h]
\begin{tabular}{cc}
\resizebox{18.8pc}{15.7pc}{\includegraphics{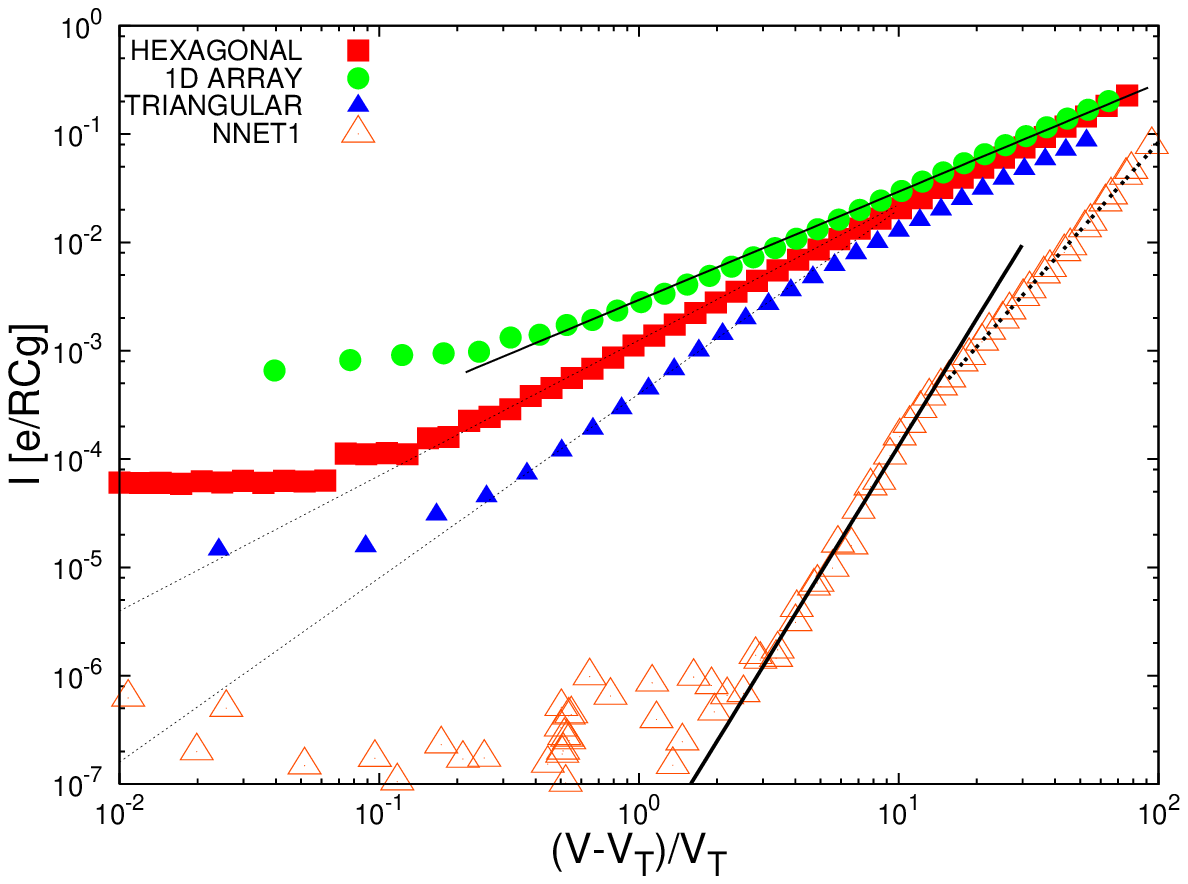}}&
\resizebox{18.8pc}{15.7pc}{\includegraphics{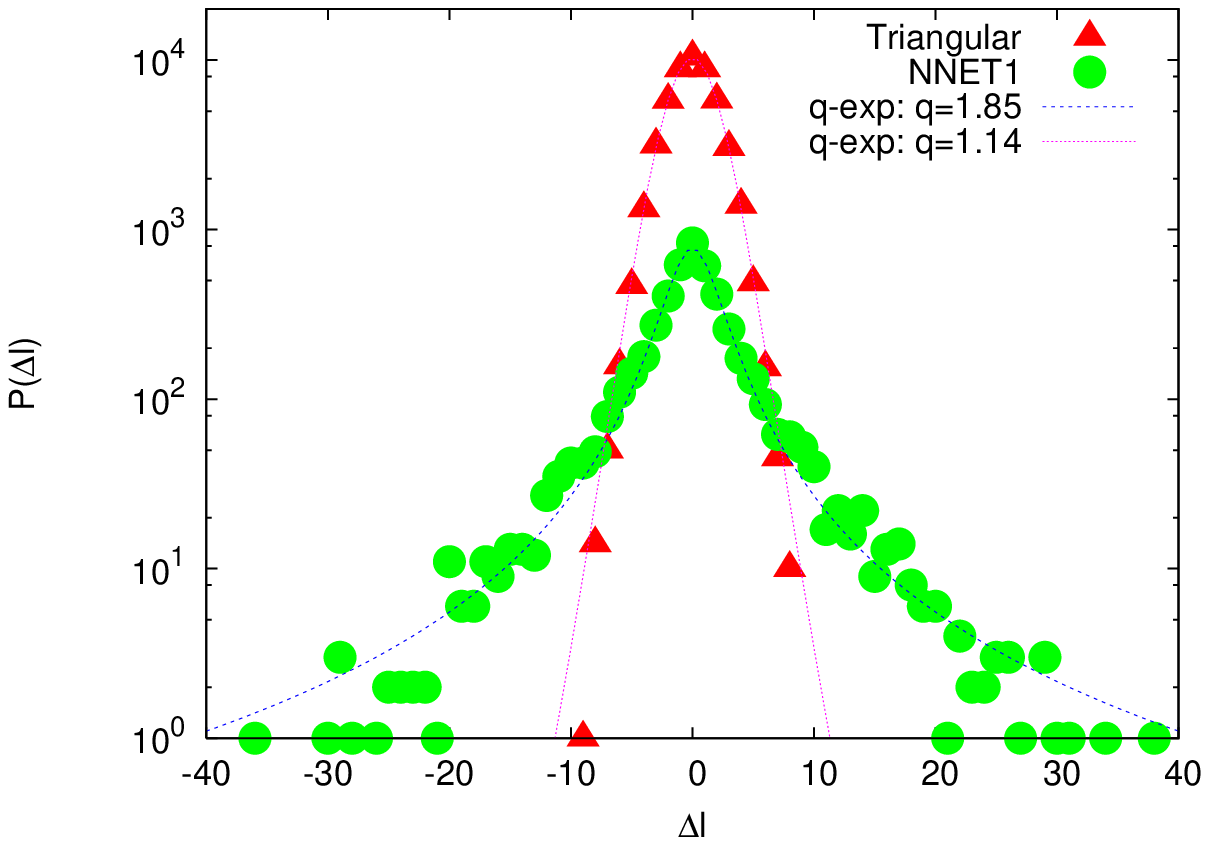}}\\
\end{tabular}
\caption{(left) Current plotted against reduced voltage for the topologically inhomogeneous nano-network NNET1 (bottom curve) and for several regular nanoparticle arrays, as indicated: triangular array with charge disorder, and hexagonal and 1-dimensional arrays. (right) Distributions of the increments of charge-fluctuations $P(\Delta I)$ plotted against $\Delta I$ for the regular triangular array with charge disorder and for the topologically disordered NNET1. Fit are $q$-Gaussian according to Eq.\ (\ref{q-Gauss}).}
\label{fig-current}
\end{figure}

In Fig.\ \ref{fig-current}(left) the current at the electrode is shown against reduced  voltage $(V-V_T)/V_T$ obtained in the simulations on different nano-networks. It is clear that for a wide range of voltages the dependences $I(V)$ are nonlinear with $\zeta >1$ according to Eq.\ (\ref{eq-IV-zeta}) for all 2-dimensional nano-networks. We have also shown the simulations for the 1-dimensional array of nanoparticles for the comparison. For the electron tunnelings through the 1-dimensional chain of nanoparticles the exponent remains $\zeta =1$ (top curve in Fig.\ \ref{fig-current}).  However, in the 2-dimensional samples many conducting paths may open between the electrodes. Flow along these paths is a dynamical process with long-range correlations. The contributions of flow along  different paths builds up the cumulative current which is measured at the electrode. Therefore, different geometries have different patterns of the conducting paths (as also shown in Fig.\ \ref{fig-cpaths}), leading to different nonlinear effects in the current--voltage dependences. Generally, the regular structures have smaller exponent, for instance, we find $\zeta \approx 2.25$ for the triangular array with charge disorder, in very good agreement with the experiment \cite{parthasarathy04}. A bit larger exponent is found in hexagonal nano-network without charge  $\zeta \approx 2.8$, whereas in the inhomogeneous nanonetwork NNET1  we have $\zeta \approx  3.9$, within the numerical error. The exponents in this range are found in other inhomogeneous nanoparticle films both in the simulations and in the experiment \cite{NL07}. 

It is interesting that such cooperative effects of different paths occurs in a limited range of voltages $V>V_T$, approximately one order of magnitude on the reduced voltage scale (see Fig.\ \ref{fig-current}). In this range of voltages, the draining of the electrons along a conducting path is intermittent and induces an avalanche of flows in the connected area of the network. However, when the voltage is large enough, a permanent flow along each paths sets in, making them chain-like, and consequently the $I(V)$ curve exhibits the crossover towards the linear dependence. 

\subsection{Time-series Analysis}
In complex dynamical systems, time series of fluctuating variables contain valuable information about the correlations in the dynamical process \cite{TT}. In the case of the single-electron transport through the nanoparticle films, we consider temporal fluctuations of the current, which is measured at the electrode, and also the fluctuations of the charges (number of tunnelings) between all nanoparticles inside the sample.

\begin{figure}[h]
\begin{tabular}{cc} 
\\\resizebox{40.8pc}{16pc}{\includegraphics{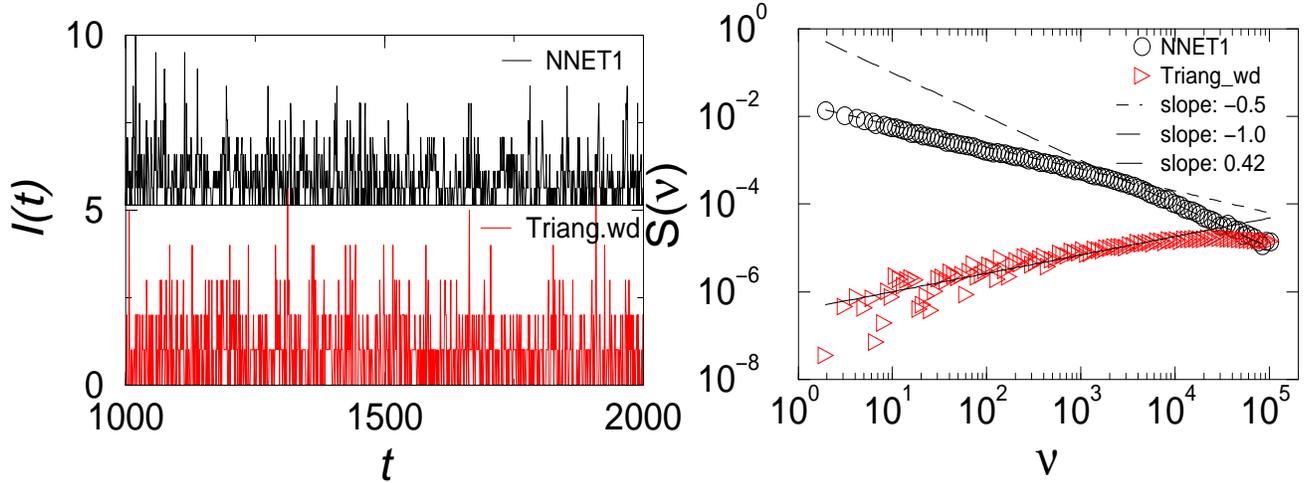}}\\
\end{tabular}
\caption{Current fluctuations time series for triangular nanoparticle array with charge disorder and for NNET1 (left) and the corresponding power spectra (right).}
\label{fig-It}
\end{figure}

For the current, which is measured at the electrode, we consider the time-series of the current $\{I(t)\}$, given in Fig.\ \ref{fig-It},  and the increments of the current $\{\Delta I(t)\}$, which is a measure of the acceleration in the electron transport along the conducting paths. Each point is averaged over time window $t_{WIN} =100$ time steps (updates of the whole network).
In Fig.\ \ref{fig-current}(right) we have shown the distributions of the increments $P(\Delta I)$ obtained for a fixed voltage above the threshold, for the triangular array with charge disorder and for the topologically disordered nanoparticle array NNET1. The distributions are fitted with the $q$-Gaussian \cite{Tsallis,Catagna}
 \begin{equation}
P(\Delta I) = B\left(1-(1-q)\left({{\Delta I}\over{D_0}}\right)^2\right)^{1/1-q}
\ ;
\label{q-Gauss}
\end{equation} 
with different $q$ values for different geometries of the nanoparticle arrays:  $q=1.14$ for the  geometrically regular triangular array with charge disorder, while $q = 1.85$  for the topologically inhomogeneous NNET1. Similarly, the analysis of the time series of the current itself, $\{I(t)\}$, exhibits different correlations in these two geometries. In Fig.\ \ref{fig-It}  the power-spectra of the time series are shown. The spectrum of the topologically inhomogeneous NNET1 shows the long-range correlations as  $S(\nu) \sim 1/\nu$ at high frequencies, whereas the spectrum flattens at high frequencies in the case of the triangular charge-disordered array. At low frequencies, however, weak correlations are observed in both networks.   

\begin{figure}[htb]
\begin{tabular}{cc} 
\resizebox{20.4pc}{!}{\includegraphics{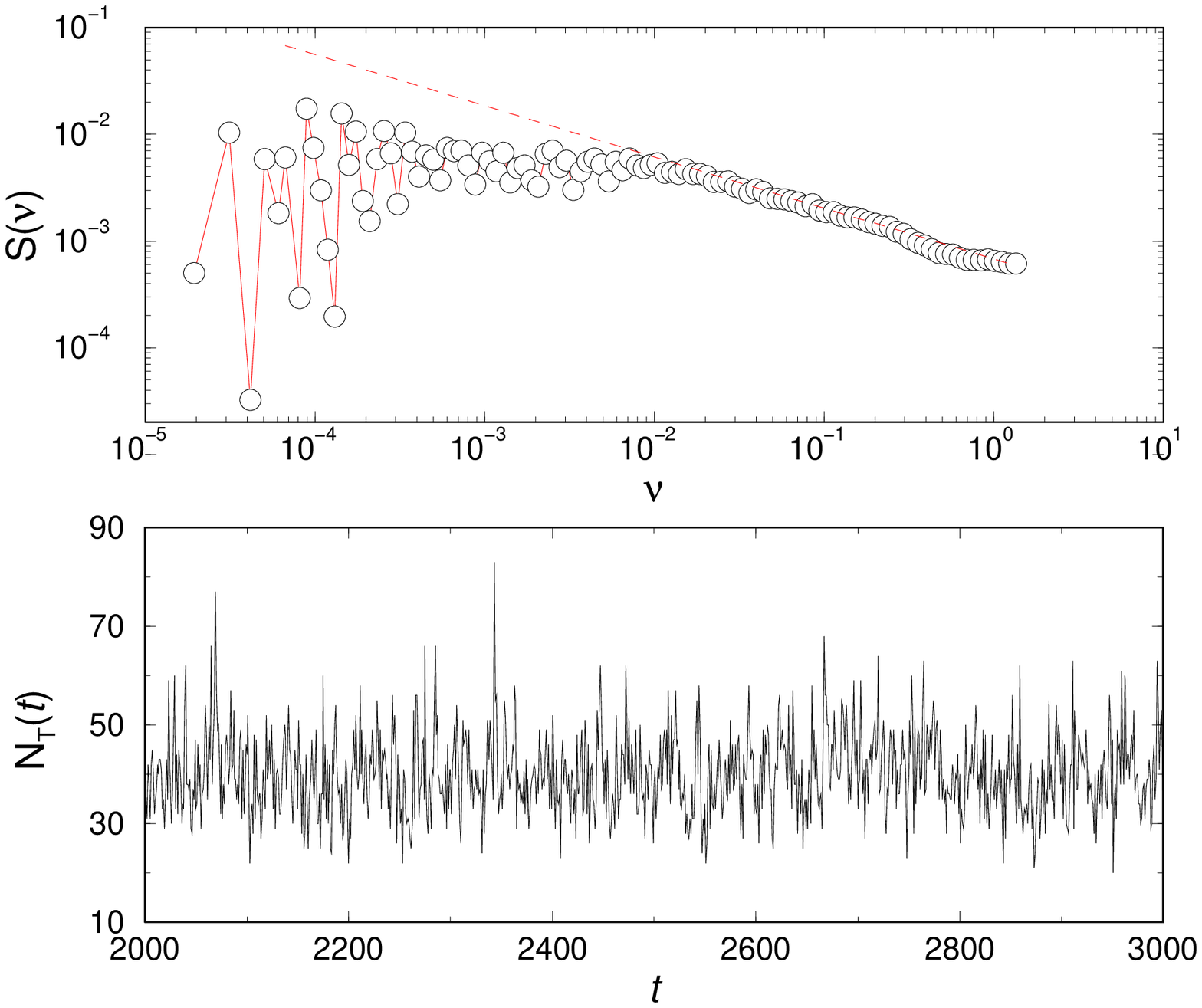}}&
\resizebox{20.4pc}{!}{\includegraphics{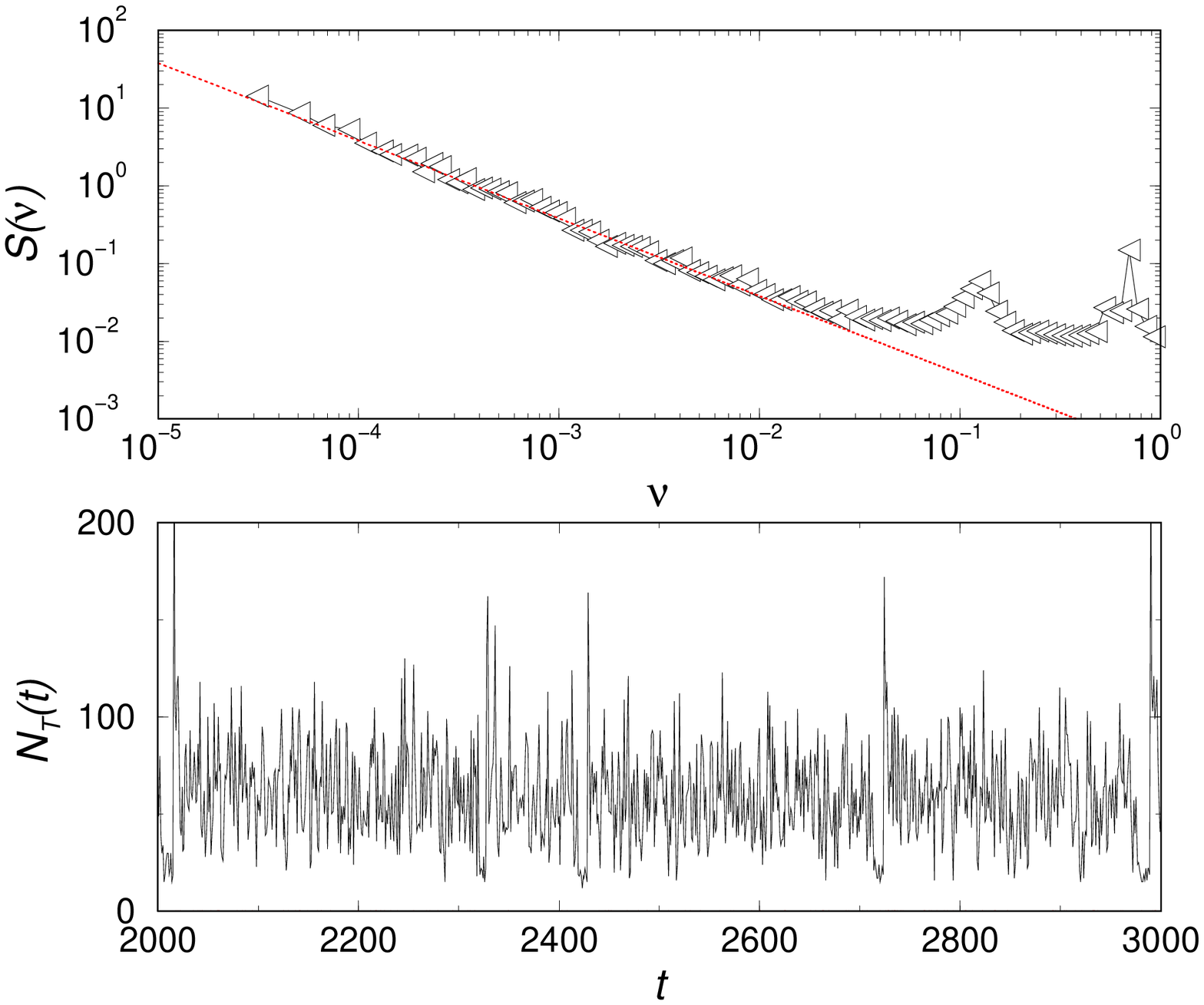}}\\
\end{tabular}
\caption{Time series representing the number of tunnelings and their corresponding power spectra in (left) the triangular array with charge disorder and (right) in the  topologically inhomogeneous NNET1 (data from Ref. \cite{ST08}).}
\label{fig-Nt}
\end{figure}
The correlations in the current fluctuations at the electrode, i.e., at the last layer of the nanoparticles connected to the electrode, are caused by the tunneling processes deeper in the sample. The time-series of the  number of tunnelings {\{$N_T(t)$\}} at all nanoparticles in in the whole network and their power spectra are shown in Fig.\ \ref{fig-Nt}. The power spectrum of the number of tunnelings exhibits long-range correlations for high frequencies, according to
\begin{equation}
S(\nu) \sim \nu^{-\phi} \ ; 
\label{eq-ps}
\end{equation}
The exponent $\phi =1$ (within numerical error bars) for the topologically disordered nano-network NNET1, suggesting strong cooperation between the conducting paths. Whereas, the correlations are much weaker in the triangular array, leading to $\phi \approx 0.3$ for high frequencies, and the absence of correlations at low frequencies. 

\section{CONCLUSIONS}
In our numerical study of the conducting nanoparticle films on substrates, we pointed out several steps that play an important role in designing the conduction properties of the films. Specifically, we study in quantitative details the emergent properties of such nanoparticle assemblies along the line from the self-assembly processes,  the topology of structures, and conduction through such structures.    
\begin{itemize}
\item{}{ \it Self-Assembly }processes are leading to different types of structures of the film. We have shown that both regular and highly inhomogeneous nanoparticle films, which are typically fabricated in entirely different techniques\cite{parthasarathy01,Mathias},  can be assembled within the same process with the bio-recognition binding and varying a single control parameter.     

\item{}{ \it Conducting paths } through the film have been identified and their topological centrality and the dynamical flow computed for for $V\geq V_T$.   Coalescence of flow along different paths  contributes to the nonlinear  $I(V)$ dependences in a range of voltages above the transition $V_T$. In our simulations, the nonlinearity exponent $\zeta$  depends strongly on the structure, in agreement with  measured  $I--V$ curves in different nanoparticle films \cite{parthasarathy01,parthasarathy04,philip02,NL07}. 

\item{}{ \it Collective dynamical effects} in the tunneling processes deep inside the sample lead to the non-Gaussian current fluctuations at the electrode. We have determined various statistical measures of these collective effects. In particular, we find  the long-range 
correlations in  charge fluctuations at all nanoparticles, 
broad distribution of flow on junctions through the nano-network, and non-Gaussian distribution of the current increments on the electrode. A non-Gaussian distribution  have been found in the experiments with nanowires \cite{Jap-wires}. Our numerical results suggest that $q-$-Gaussian distribution with different parameters could be measured in the nanoparticle films. The occurrence of the $q-$Gaussian in a larger class of complex dynamical systems with long-range interactions has been discussed recently \cite{Catagna1}.   
\end{itemize}
Furthermore, we demonstrate how different types of disorder affect the conduction in our comparative study of the regular nanoparticle array in the presence of local charge disorder, on one side, and the  topologically inhomogeneous nano-network without charge disorder, on the other.  
We hope that our detailed numerical study of the processes leading to different structures and the structure-dependent function of the nanoparticle assemblies sheds a new light on the problem of engineering of the functional materials with desired characteristics.
\acknowledgments
We thank the project MRTN-CT-2004-005728 (EC) and the program P1-0044 (Slovenia) for the support.

\end{document}